\documentclass[letterpaper]{jpconf}
\usepackage{graphicx}
\begin{document}
\title{Inclusive distributions of 
charged hadrons in pp collisions at $\sqrt{s}$ = 0.9 and 2.36~TeV}

\author{G\'abor I. Veres on behalf of the CMS Collaboration}

\address{CERN, Geneva\footnote{On leave from 
E\"otv\"os Lor\'and University, Budapest}}

\begin{abstract} 
Measurements of inclusive charged-hadron transverse-momentum ($p_T$) and 
pseudorapidity ($\eta$) distributions are presented for proton-proton 
collisions at $\sqrt{s}=$~0.9 and 2.36~TeV. 
For non-single-diffractive interactions, the average $p_T$ of
charged hadrons is measured to be 
$0.46 \pm 0.01$~(stat.)~$\pm$~0.01~(syst.)~GeV/$c$ at 0.9~TeV and 
$0.50 \pm 0.01$~(stat.)~$\pm$~0.01~(syst.)~GeV/$c$ at 2.36~TeV, for 
$|\eta|<2.4$. At these energies, the measured pseudorapidity densities 
in the central region, $dN_{\rm ch}/d\eta|_{|\eta| < 0.5}$, are 
$3.48 \pm 0.02$~(stat.)~$\pm$~0.13~(syst.) and 
$4.47 \pm 0.04$~(stat.)~$\pm$~0.16~(syst.), respectively.
The results at 2.36~TeV represent the highest-energy 
measurements ever published at a particle collider at the time of 
the presentation at the Lake Louise Winter Institute. 
\end{abstract}

\vspace{-4.4mm}
\section{Introduction} 

Measurements of particle yields and kinematic distributions are an 
essential first step in exploring a new energy regime of particle 
collisions. Such studies contribute to our understanding of the physics 
of hadron production and help construct a solid foundation for 
other investigations. In the complicated environment of LHC $pp$ 
collisions~\cite{Evans:2008zzb}, such studies are needed to distinguish 
rare signal events from the much larger backgrounds of soft hadronic 
interactions. They will also serve as points of reference for
Pb-Pb collisions in the LHC.
Soft collisions are commonly classified as elastic scattering, inelastic 
single-diffractive (SD) dissociation, double-diffractive (DD) 
dissociation, and inelastic non-diffractive (ND) 
scattering~\cite{Kittel:2005fu}. All presented results refer to 
inelastic non-single-diffractive (NSD) interactions.
Primary charged hadrons are defined to include decay products of 
particles with proper lifetimes less than 1~cm. Products of secondary 
interactions and leptons are excluded. The observables reported here 
are $dN_{\rm ch}/d\eta$ and $dN_{\rm ch}/dp_{\rm T}$
in the $|\eta| < 2.4$ range~\cite{1stpaper}.
The data for this study were recorded with the Compact Muon Solenoid 
(CMS) experiment~\cite{JINST} in December 2009, during a few hours of 
the early LHC operation at $\sqrt{s}=0.9$ and 2.36~TeV. 

\section{Experimental methods} 

A detailed description of the CMS experiment can be found in 
Ref.~\cite{JINST}. The detectors used for the present analysis are the 
pixel and silicon-strip tracker (SST), covering the region $|\eta|< 2.5$ 
and immersed in a 3.8~T axial magnetic field. The pixel tracker consists 
of three barrel layers and two end-cap disks at each barrel end. The 
forward calorimeter (HF), which covers the region $2.9<|\eta|<5.2$, was 
also used for event selection. The detailed Monte Carlo simulation (MC) 
of the CMS detector response is based on GEANT4 \cite{GEANT4}. The 
inelastic $pp$ collision rate was about 3-11~Hz. The fraction of 
the events in the data with more than one collision was less than 
$2\times 10^{-4}$ and was neglected. Any hit in the beam scintillator 
counters (BSC, $3.23<|\eta|<4.65$) coinciding with colliding proton 
bunches was used for triggering the data acquisition. In addition,
a reconstructed primary vertex (PV) was required using the tracker,
together with at least one HF tower in each end with more than 3~GeV 
total energy. The fraction of beam-halo and other beam-background 
events were suppressed below $0.1\%$. The number of selected events was 
finally 40,320 and 10,837 at 0.9 and 2.36~TeV, respectively.

The event selection efficiency was estimated with simulated events using 
the PYTHIA~\cite{Sjostrand:2006za} and 
PHOJET~\cite{Bopp:1998rc, Engel:1995sb} event generators. The relative 
event fractions of SD, DD, and ND processes and their respective event 
selection efficiencies are different for the two models, but the 
estimated fractions of SD events in the selected data samples are 
similar: 5.2\% (4.9\%) at 0.9~TeV and 6.3\% (5.0\%) at 2.36~TeV for 
PYTHIA (PHOJET), respectively. The selection efficiency of NSD processes 
as a function of multiplicity, and the above fraction of SD 
events are corrected for.

The $dN_{\rm ch}/d\eta$ distributions were obtained
with three methods, based on counting the following quantities: (i) 
reconstructed clusters in the barrel part of the pixel detector; (ii) 
pixel tracklets composed of pairs of clusters in different pixel barrel 
layers; and (iii) tracks reconstructed in the full tracker volume. The 
third method also allows a measurement of the 
$dN_{\rm ch}/dp_{\rm T}$ distribution. All these methods rely on the 
reconstruction of a PV~\cite{Sikler:2009nx}. The three methods are sensitive 
to the measurement of particles down to $p_T$ values of about 30, 50, 
and 100~MeV/$c$, respectively.

The measurements were corrected for the geometrical acceptance, 
efficiency, fake and duplicate tracks, low-$p_T$ particles curling in the 
axial magnetic field, decay products of long-lived hadrons and photon 
conversions and inelastic hadronic interactions in the detector 
material. The PYTHIA D6T tune was chosen to determine the corrections.

\section{Results} 

\begin{figure}
 \begin{flushleft}
  \includegraphics[width=0.48\textwidth,height=0.5\textwidth]{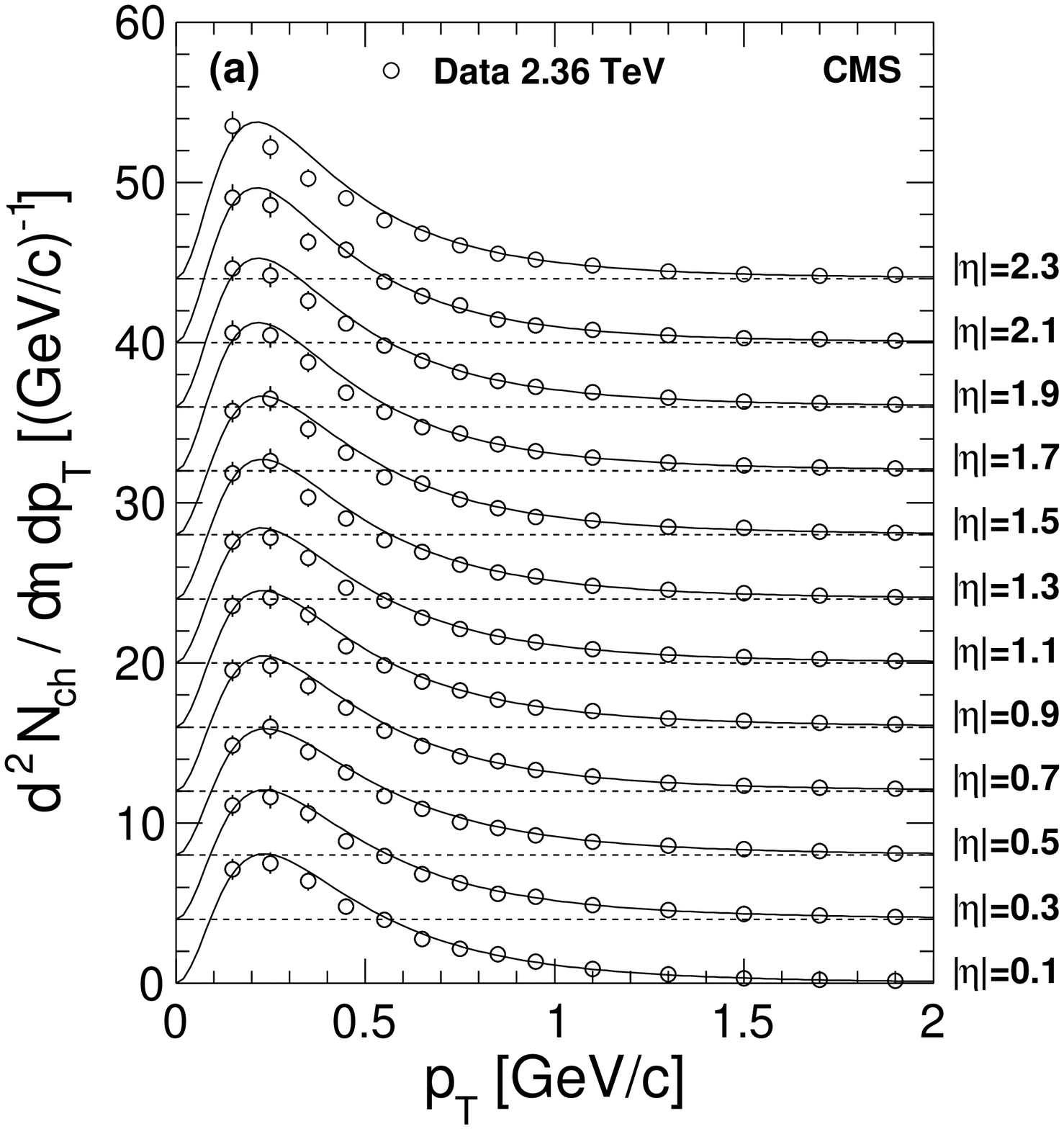}
  \hspace{0cm}
  \includegraphics[width=0.48\textwidth,height=0.5\textwidth]{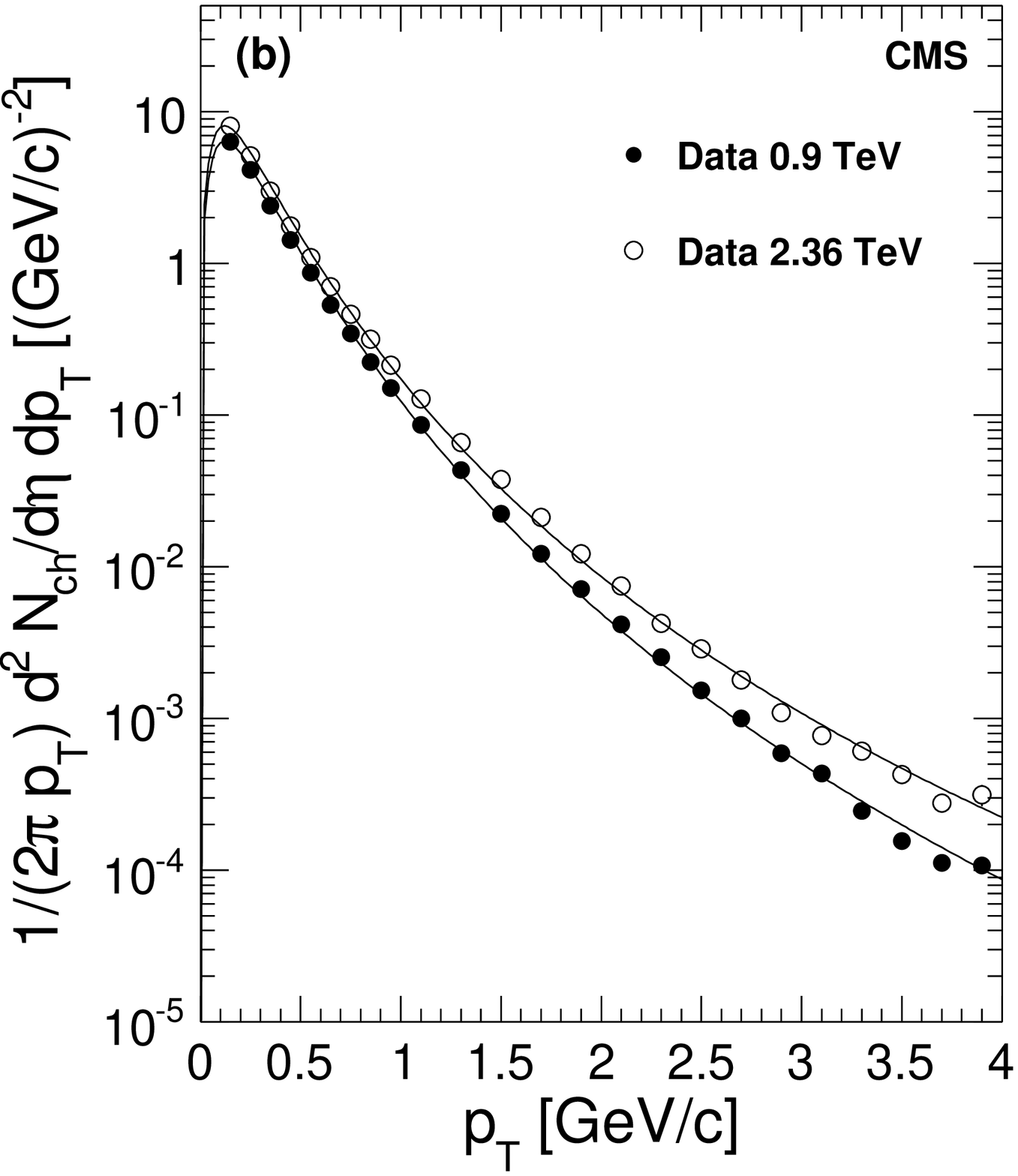}
 \end{flushleft}
\vspace{-3mm}
 \caption{
(a) Measured differential yield of charged hadrons in the range $|\eta| 
< 2.4$ in $0.2$-unit-wide bins of $|\eta|$ for the 2.36~TeV data. The 
measured values with systematic uncertainties (symbols) and the fit 
functions (Eq.~\ref{eq:tsallis}) are displayed. The values with 
increasing $\eta$ are successively shifted by four units along the 
vertical axis.
(b) Measured yields of charged hadrons for $|\eta| < 2.4$ with systematic
uncertainties (symbols), fitted with the empirical function
(Eq.~\ref{eq:tsallis}).}
 \label{fig:spectra}
\end{figure}

For the measurement of the $p_{\rm T}$ distribution, 
charged-particle tracks with $p_T>0.1$~GeV/$c$ were used in 
12 different $|\eta|$ bins, from 0 to 2.4. The average charged-hadron 
yields in NSD events are shown in Fig.~\ref{fig:spectra}a as a function 
of $p_T$ and $|\eta|$. The Tsallis parametrization 
\cite{Tsallis:1987eu,Wilk:2008ue,Biro:2008hz},

\begin{equation}
 E \frac{d^3N_{\mathrm{ch}}}{d p^3} =
\frac{1}{2\pi p_T} \frac{E}{p} \frac{d^2N_{\mathrm{ch}}}{d\eta\,dp_T} = 
C \frac{dN_{\mathrm{ch}}}{dy}\left(1 + 
\frac{E_\mathrm{T}}{nT}\right)^{-n},
 \label{eq:tsallis}  
\end{equation}

\noindent was fitted to the data. The $p_T$ spectrum of charged 
hadrons measured in the range $|\eta| < 2.4$, is shown in 
Fig.~\ref{fig:spectra}b for data at 0.9 and 2.36~TeV.
The average $p_T$, calculated from a combination of the 
measured data points and the low- and high-$p_T$ contributions as 
determined from the fit, is $\langle p_T \rangle = 0.46 \pm 
0.01 (stat.)\pm 0.01 (syst.)$~GeV/$c$ and 
$0.50 \pm 0.01 (stat.)\pm 0.01 (syst.)$~GeV/$c$ at 0.9 and 2.36~TeV, 
respectively.

Experimental uncertainties related to the trigger and event selection 
are common to all the analysis methods. The uncertainty related to the 
presence of SD and DD events in the final sample was estimated to 
be 2\%, based on consistency checks between data and simulation for 
diffractive event candidates. The total event selection uncertainty, 
which also includes the selection efficiency of the BSC and HF, was 
found to be 3\%. Additional 3\% and 2\% uncertainties were assigned 
to the tracklet and track reconstruction algorithm efficiencies, 
respectively. Corrections at the percent level were applied to the final 
results to extrapolate to $p_T = 0$. The uncertainty on these 
extrapolation corrections was found to be less than 1\%. 
The final systematic uncertainties for the pixel counting, tracklet, 
and track methods were found to be 5.4\%, 4.9\%, and 4.0\%, 
respectively, and are strongly correlated.

\begin{figure}[h]
 \begin{center}
  \includegraphics[width=0.48\textwidth]{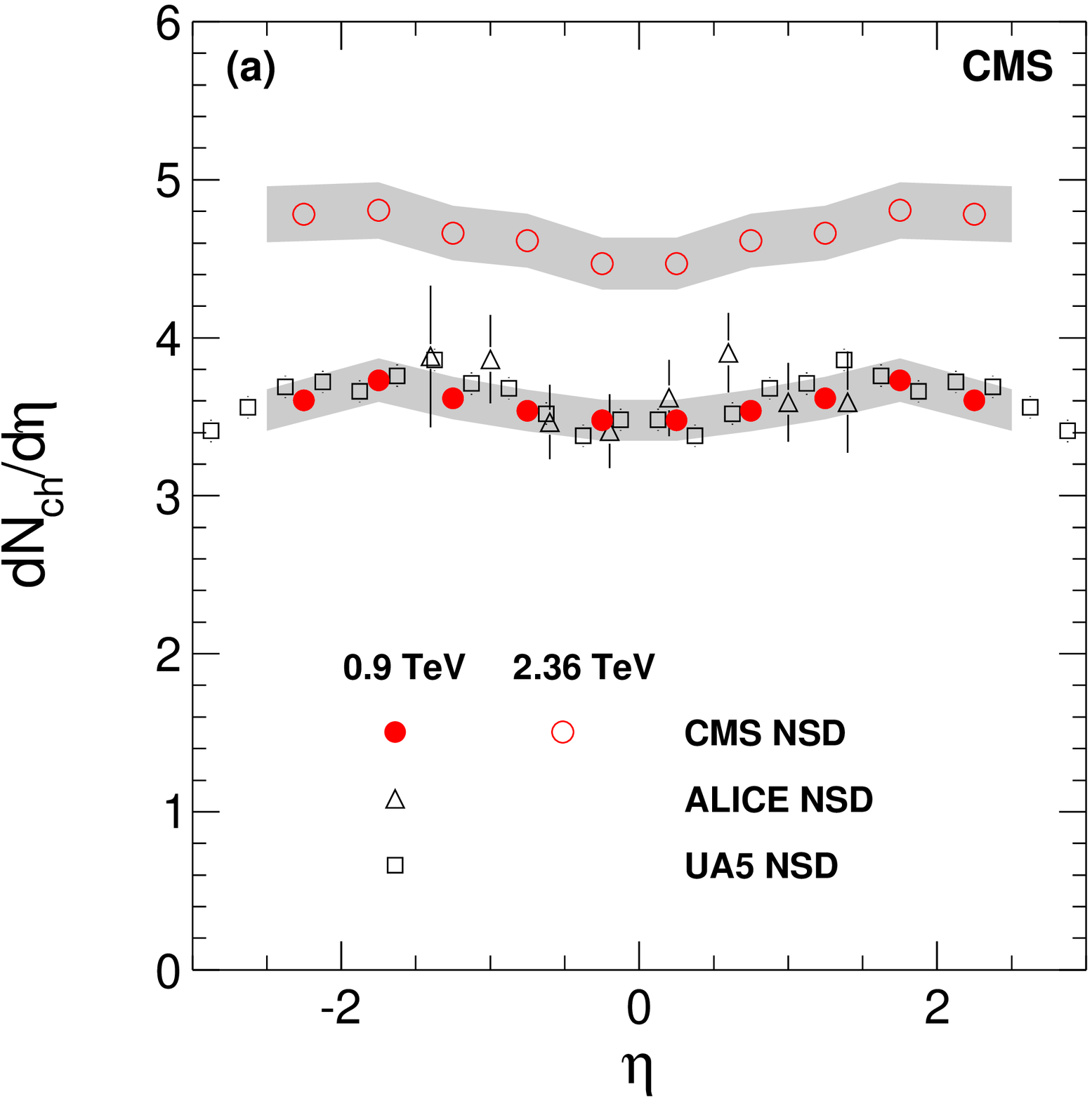}
  \hspace{0cm}
  \includegraphics[width=0.48\textwidth]{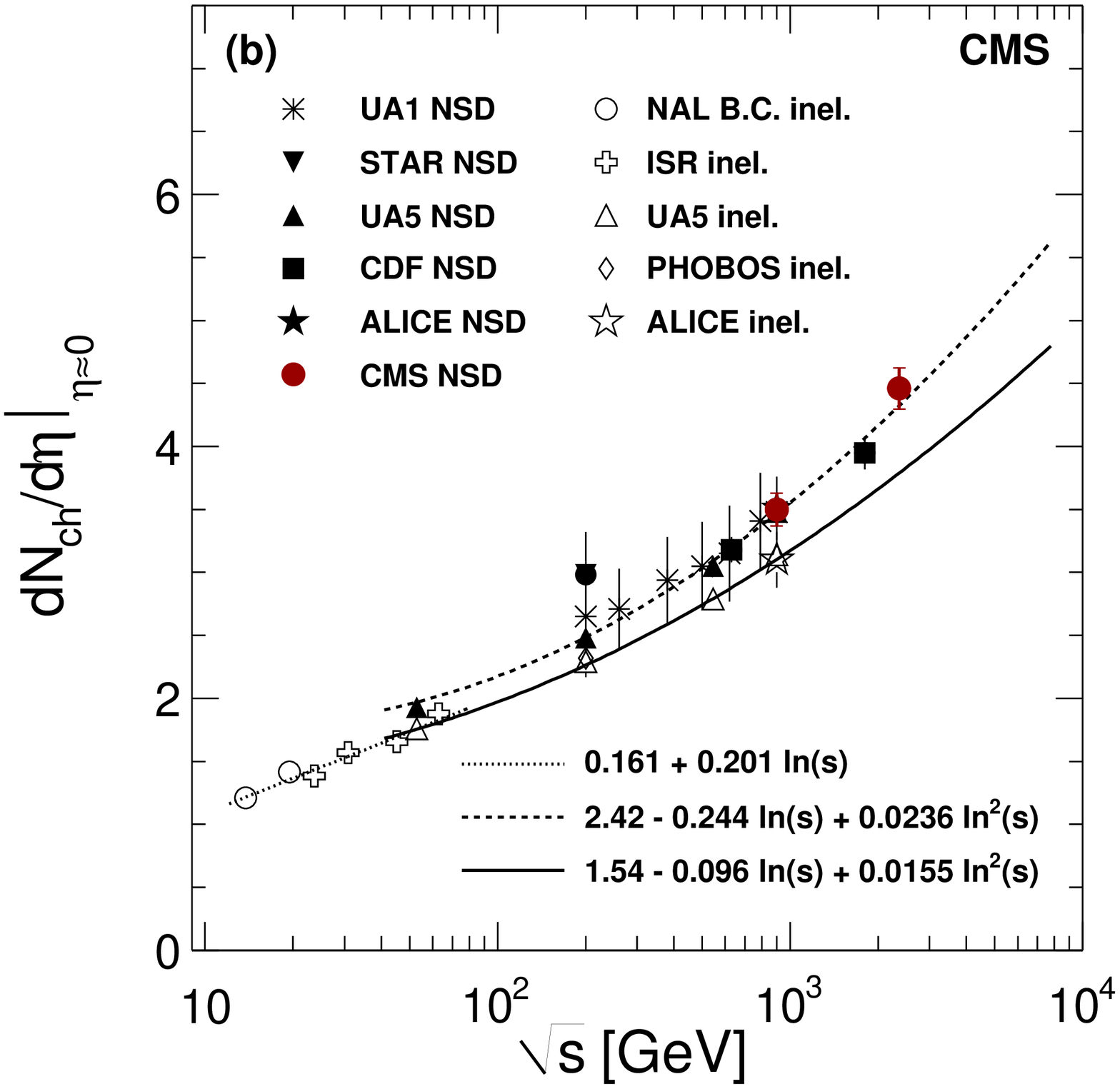}
 \end{center}
 \caption{
(a) Reconstructed $dN_{\rm ch}/d\eta$ distributions averaged over the 
cluster counting, tracklet and tracking methods (circles), compared to 
data from the UA5 (open squares) and from the ALICE (open triangles) 
experiments at 0.9~TeV, and the averaged result over the three methods 
at 2.36~TeV (open circles). The CMS and UA5 data points are symmetrized 
in $\eta$. The shaded band represents systematic uncertainties of this 
measurement, which are largely correlated point-to-point. The error bars 
on the UA5 and ALICE data points are statistical only.
(b) Charged-hadron pseudorapidity density in the central region as a 
function of centre-of-mass energy in $pp$ and $p\overline{p}$ collisions 
including lower energy data, together with various empirical 
parameterizations fit to the data corresponding to the inelastic (solid 
and dotted curves with open symbols) and to the NSD (dashed curve with 
solid symbols) event selection. The error bars indicate systematic 
uncertainties, when available.}
\label{fig:dndeta}
\end{figure}

For the $dN_{\rm ch}/d\eta$ measurements, the results from the three 
different methods were averaged. The final $dN_{\rm ch}/d\eta$ 
distributions are shown in Fig.~\ref{fig:dndeta}a for $\sqrt{s}=0.9$
and 2.36~TeV. 
For $|\eta| < 0.5$, the average charged multiplicity density is   
$3.48 \pm 0.02$~(stat.)~$\pm$~0.13~(syst.) and 
$4.47 \pm 0.04$~(stat.)~$\pm$~0.16~(syst.) for NSD events at 0.9 and 
2.36~TeV, respectively. The $\sqrt{s}$ dependence of the measured 
$dN_{\rm ch}/d\eta|_{\eta \approx 0}$
is shown in Fig.~\ref{fig:dndeta}b, which includes data from 
the NAL Bubble Chamber~\cite{Whitmore:1973ri}, the 
ISR~\cite{Thome:1977ky}, and UA1~\cite{Albajar:1989an}, 
UA5~\cite{Alner:1986xu}, CDF~\cite{Abe:1989td}, STAR~\cite{star:2008ez}, 
PHOBOS~\cite{Nouicer:2004ke} and ALICE~\cite{alice}.
The $dN_{\rm ch}/d\eta$ results reported here 
show a rather steep increase between 0.9 and 2.36~TeV, which is measured 
to be ($28.4 \pm 1.4\, {\rm (stat.)} \pm 2.6\, \rm{(syst.)}$)\%, 
significantly larger than the 18.5\% (14.5\%) increase predicted by 
PYTHIA (PHOJET).

In summary, charged-hadron $p_T$ and $\eta$ distributions have 
been measured in proton-proton collisions at $\sqrt{s} = 0.9$ and 
2.36~TeV. The measured increase of the pseudorapidity density
between these energies is higher than most predictions and provides new 
information to constrain ongoing improvements of soft particle 
production models and event generators. 
The mean $p_T$ of charged hadrons has been also measured, extrapolated 
to the full $p_T$ range. These studies are the first steps in 
the exploration of particle production at the new centre-of-mass energy 
frontier, and contribute to the understanding of the dynamics in soft 
hadronic interactions.

\noindent 
{\bf Acknowledgements.} We congratulate and express our 
gratitude to our colleagues in the CERN accelerator departments for the 
excellent performance of the LHC. We thank the technical and 
administrative staff at CERN and other CMS institutes, and acknowledge 
support from: FMSR (Austria); FNRS and FWO (Belgium); CNPq, CAPES, 
FAPERJ, and FAPESP (Brazil); MES (Bulgaria); CERN; CAS, MoST, and NSFC 
(China); COLCIENCIAS (Colombia); MSES (Croatia); RPF (Cyprus); Academy 
of Sciences and NICPB (Estonia); Academy of Finland, ME, and HIP 
(Finland); CEA and CNRS/IN2P3 (France); BMBF, DFG, and HGF (Germany); 
GSRT (Greece); OTKA and NKTH (Hungary); DAE and DST (India); IPM (Iran); 
SFI (Ireland); INFN (Italy); NRF and WCU (Korea); LAS (Lithuania); 
CINVESTAV, CONACYT, SEP, and UASLP-FAI (Mexico); PAEC (Pakistan); SCSR 
(Poland); FCT (Portugal); JINR (Armenia, Belarus, Georgia, Ukraine, 
Uzbekistan); MST and MAE (Russia); MSTDS (Serbia); MICINN and CPAN 
(Spain); Swiss Funding Agencies (Switzerland); NSC (Taipei); TUBITAK and 
TAEK (Turkey); STFC (United Kingdom); DOE and NSF (USA).
The author thanks the Hungarian Scientific Research Fund
(H07-C74248, K81614 and NK81447) for its support.

\end{document}